\documentclass[aps,prl,epsf,floats,epsfig,aps,amsmath,amssymb,showpacs]{revtex4}

\input{epsf}
\begin{document}
\def\limo{\mathrel{\mathop{\longrightarrow}\limits^{{\cal P}}}}
  
\draft 

\title
{Spin network quantum simulator}
\author {Annalisa Marzuoli$^{\divideontimes\, ,\ddagger}$, and Mario Rasetti$^{\#\, 
,\circledast}$}
\address{$^{\divideontimes}$Dipartimento di Fisica Nucleare e Teorica, Universit\`a 
di Pavia, I-27100 Pavia, Italy \\ $^{\ddagger}$Istituto Nazionale di Fisica Nucleare, 
Sezione di Pavia, I-27100 Pavia, Italy \\ $^{\#}$Dipartimento di Fisica, Politecnico 
di Torino, I-10129 Torino, Italy \\ $^{\circledast}$Istituto Nazionale di Fisica della 
Materia, Unit\`a Politecnico di Torino, I-10129 Torino, Italy}

\begin{abstract}
{We propose a general setting for a universal representation of the quantum structure on 
which quantum information stands, whose dynamical evolution (information manipulation) is 
based on angular momentum recoupling theory. Such scheme complies with the notion of 'quantum 
simulator' in the sense of Feynman, and is shown to be related with the topological quantum 
field theoretical approach to quantum computation. }
\end{abstract}

\maketitle


\bigskip
\section{Introduction}

In the past few years there has been a tumultuous activity aimed to proposing 
novel conceptual schemes of interpretation of quantum computation. Curiously 
enough, most of them are based on topological notions. Among these, anyonic 
quantum computation \cite{Kit2}, fermionic quantum computation \cite{Kit1}, 
localized modular functor quantum field computation \cite{FrLaWa}, holonomic 
quantum computation \cite{ZaRa} have mostly attracted attention. Such models 
appear to be simply different realizations of a unique conceptual scheme that 
incorporates all of them as particular instances. In this note we aim to claiming 
that such schemes may all be identified with a model of quantum simulator (in the 
sense of Feynman \cite{Fey}) based on (re)coupling theory of $SU(2)$ angular 
momenta (see, e.g. \cite{BiLo} and references therein). The paper raises a number 
of issues without entering in too many technical details but rather trying to 
establish the guiding philosophy, and is therefore foundational. 

The scheme automatically incorporates all the essential features that make quantum 
information encoding so much more efficient than classical: it is fully discrete 
(including its time-like variable); it deals with inherently entangled states, and 
thus incorporates all achievable complexity in its set-up, which is naturally 
endowed with a tensor product  structure; it allows for generic encoding patterns. 
  
In ref.\cite{Fey} Feynman lists a minimal set of requirements as essential for 
the proper characterization of an efficient quantum simulator: i) locality of 
interactions; ii) number of 'computer' elements proportional to (a function at 
most polynomial of) the space-time volume of the physical system; iii) time 
discreteness (time is 'simulated' in the computer by computational steps). Our 
argument is based on the fact that all such basic features are typical of spin 
networks. 

By 'spin networks' we mean here -- contrary to what happens in solid state physics, 
but somewhat in the spirit of combinatorial approach to space-time representation 
\cite{Pen} -- graphs the node and edge sets of which can be labelled by quantum 
numbers associated with $SU(2)$ irreducible representations and by $SU(2)$ recoupling 
coefficients, respectively. Spin networks can thus be thought of as an ideal candidate 
conceptual framework for dealing with tensorial transformations and topological effects 
in groups of observables. The idea is to exploit to their full extent the discreteness 
hypotheses ii) and iii), by modelling the computational space in terms of a set of 
combinatorial and topological rules that mimic space-time features in a way that 
automatically includes quantum mechanics. 

\bigskip  
\section{The Simulator Computational Space}

We begin by defining the structural setting of a universal quantum simulator 
${\mathfrak{M}}$ which satisfies all axioms proper to the quantum Turing Machine 
\cite{Man}.  ${\mathfrak{M}}$, whose computational space is identified with 
a spin network, can encode information, undergo unitary transformations, and 
simulate any finite quantum system completely described by eigenstates of $SU(2)$ 
angular momentum operators. In the sequel we shall indicate how such a scheme can 
be extended to mixed states. 

\noindent{\bf Coding Information} 

The machine ${\mathfrak{M}}$ building blocks are an ordered collection of $n+1$ mutually 
commuting angular momentum operators $\{ {\bf J}_{\ell} \, |\, \ell =1,\dots , n+1 \}$ (for 
example associated with a set of $n+1$ kinematically independent particles), with eigenvalues 
parametrized by $j_1,\dots ,j_{n+1}$, with $j_{\ell} = 0,\frac{1}{2},1, \frac{3}{2},\dots $. 
Such operators are assumed to sum to a sharp total angular momentum ${\bf J}$ with 
projection $J_z$, whose quantum numbers are, respectively, $j$ and $m$, $-j \leq m \leq 
j$ in integer steps. 
 
For any given pair $(n,j)$, all possible binary coupling schemes of the $n+1$ angular momenta 
$j_{\ell}$ together with the the quantum numbers $k_1, \dots , k_{n-1}$ corresponding to the $n-1$ 
intermediate angular momenta, provide the 'alphabet' in which quantum information is encoded. 
The resulting Hilbert spaces ${\cal H}^j_n(k_1,\dots ,k_{n-1})$, each $(2j+1)$-dimensional, are 
spanned by complete orthonormal sets of the form 
\begin{eqnarray}
\bigl \{ | j_1, \dots , j_{n+1}  \, ;\, k_1,\dots ,k_{n-1}\, ;\, j,m \rangle \equiv 
| {\mathfrak{b}} \rangle \bigr \} \; . 
\label{base} 
\end{eqnarray} 

\begin{figure}[htb]
\leavevmode
\epsfysize=8.5cm
\epsfbox{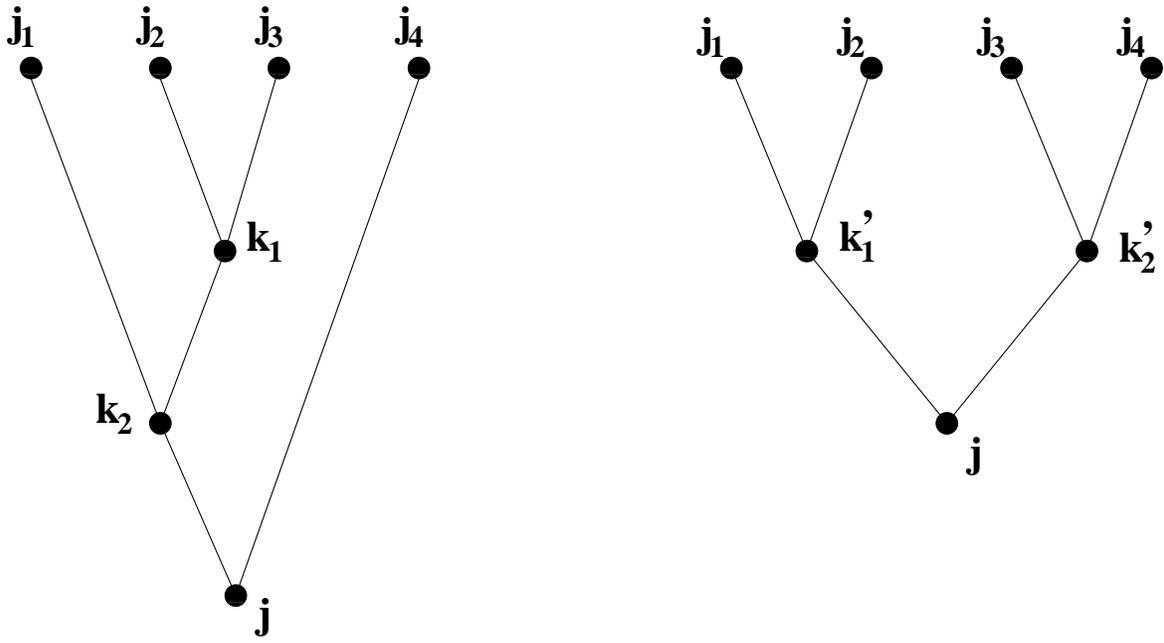}
\bigskip 
\caption{
Different rooted binary trees on $(n+1)=4$ terminal nodes 
are depicted. Their $(2n+1)=7$ nodes are labelled by angular momentum eigenvalues. 
The tree on the left corresponds to the binary bracketing  
$| j_1,j_2,j_3,j_4\, ; \, k_1,k_2 \, ;\, j,m\rangle$ $\longrightarrow$  
$\bigl ( \bigl (j_1 \bigl (j_2j_3 \bigr )_{k_1}\bigr )_{k_2}j_4 \bigr )_j$.
The tree on the right corresponds to the binary bracketing 
$| j_1,j_2,j_3,j_4\, ; \, k'_1,k'_2 \, ;\, j,m\rangle$ $\longrightarrow$    
$\bigl ( \bigl (j_1j_2 \bigr )_{k'_1} \bigl (j_3j_4 \bigr )_{k'_2} \bigr )_j$.
}
\label{binary}
\end{figure}

Such states can be pictorially represented by rooted labelled binary trees in which each node 
corresponds to an angular momentum quantum number: the root of the tree to $j$, the internal nodes 
to the intermediate $k_1,\dots ,k_{n-1}$, and the terminal nodes to $j_1,\dots ,j_{n+1}$.  
An equivalent representation is the binary bracketings notation proposed in \cite{BiLo}. Fig.\ref{binary}  
shows an example of these kinds of representation where, fixed an ordering $j_1,j_2,\ldots,j_{n+1}$ and 
given a common $j$, there exists a correspondence between states given in ({\ref{base}}) and the (equivalent)
combinatorial structures represented by binary bracketings and labelled binary trees (cfr. {\it e.g.} 
\cite{Sta}).

\begin{center}
{\bf Fig. 1}
\end{center}

Notice moreover that to each of these assignments (states (\ref{base})), there corresponds 
a unique non-associative structure over the tensor product ${\cal H}^{j_1} \otimes\cdots\otimes 
{\cal H}^{j_{n+1}} \equiv {\rm span} \{ |j_1m_1\rangle \otimes \cdots \otimes |j_{n+1}m_{n+1}\rangle 
\}$, which is manifestly intrinsically entangled. 

A code, in this picture, is a sort of generalized quantum G\"odel number in the 
base associated with a field made of the ordered labels of the intermediate angular 
momenta, which range over a finite domain ({\it e.g.} for ${\bf J}_1 + {\bf J}_2 
= {\bf J}_{12}$, $|j_1 -j_2|\leq j_{12} \leq j_1+j_2$), as well as the coupling 
brackets. Such coding spans (and defines) the space of all possible 
computational states.    
 
\noindent {\bf Operations as Unitary Transformations} 

In the structure described above, any quantum operation is implemented by some  
transformation connecting pairs of binary coupled states, namely by the 
so called 'recoupling coefficients', or $3nj$ symbols \cite{VaMoKh}, 
\cite{BiLo}. 

Indeed the $3nj$ symbols are unitary probability amplitudes 
\begin{eqnarray}
  {\cal U}_{3nj} \left [ {{k_1,\dots ,k_{n-1}}\atop{k'_1,\dots ,k'_{n-1}}} \right ] \doteq 
  \langle j_1, \dots , j_{n+1}  \, ;\, k'_1,\dots ,k'_{n-1}\, ;\, j,m \, | \, j_1, \dots 
  , j_{n+1}  \, ;\, k_1,\dots ,k_{n-1}\, ;\, j,m \rangle \; , \nonumber  
\end{eqnarray}
$|{\cal U}_{3nj}|^2$ representing the probability that the system, once prepared 
in state $| j_1, \dots , j_{n+1} ; k_1,\dots ,k_{n-1} ; j,m \rangle$, is measured 
in state $| j_1, \dots , j_{n+1} ;\, k'_1, \dots ,k'_{n-1} ; j,m \rangle$. 

Notice that the recoupling coefficients can be interpreted as reduced 
matrix elements since the total magnetic quantum number $m$ can be neglected in view of the 
Wigner-Eckart theorem. Since they give the elements of the transfer matrices connecting any 
pair of states, the symbols actually provide the (matrix) analog of the transition function 
of the quantum Turing Machine \cite{Man}. 

Moreover as any transfer from a state to another -- states being in 
one-to-one correspondence with the vertices of a suitable graph ${\mathfrak{G}}_n$ as we shall 
see below -- can be thought of as a (discrete) path integral, $3nj$ symbols implicitly define 
an 'action' (and hence an associated hamiltonian operator). 

We shall show in the next section that programming ${\mathfrak{M}}$ consists just in 
selecting which transformations do perform the desired computation.
  
\noindent {\bf Computational Space of the Machine}  

The computational space associated with ${\mathfrak{M}}$ is a graph ${\mathfrak{G}}_n$ 
whose vertices are identified ({\sl i.e.} are in one-to-one correspondence) with the system 
pure angular momentum eigenstates defined in (\ref{base}).  

\begin{figure}[htb]
\leavevmode
\hspace{+1cm}
\epsfbox{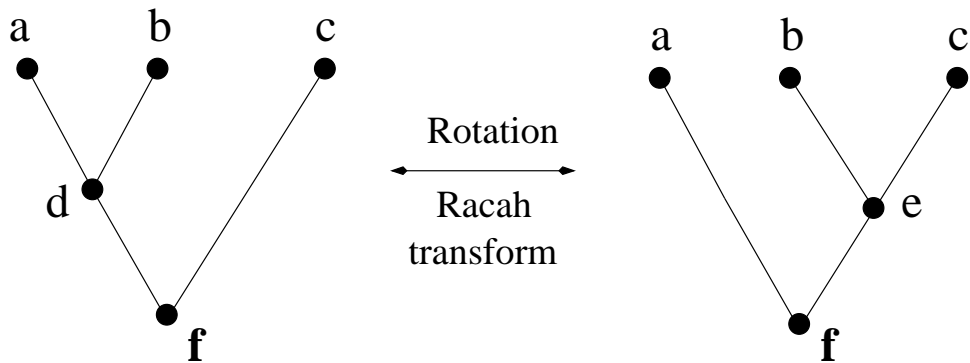}
\bigskip 
\caption{
The rotation operation on a portion of a labelled binary tree. The explicit 
expression of the Racah transform (\ref{A}) relating the states associated with the trees
depicted here reads $|(a\,(bc)_e\,)_f\,;m>\,=\, \sum_{d}$ $(-1)^{a+b+c+f}\; [(2d+1)
(2e+1)]^{1/2}$ $\bigl \{ {{a\, b\, d}\atop{c\, f\, e}} \bigr \}\;|(\,(ab)_d \,c)_f \,;m>$  
where the unitary matrix $\{6j\}$ is the Racah--Wigner $6j$ symbol of $SU(2)$.
}
\label{rotation}
\end{figure}

\begin{figure}[htb]
\leavevmode
\hspace{+2.5cm}
\epsfbox{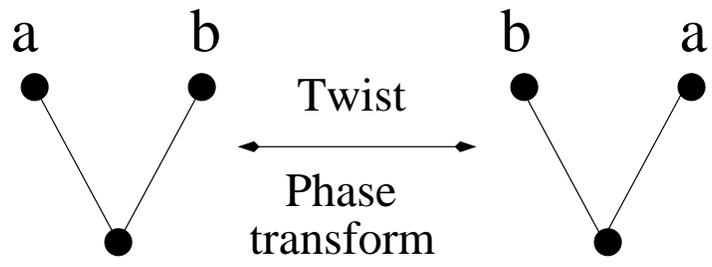}
\bigskip 
\caption{The twist operation on a portion of a labelled
binary tree. According to (\ref{B}) the quantum state changes
only by a phase factor.}
\label{twist}
\end{figure}

The Racah transform ${\cal R}$, together with the phase transform $\Phi$,  
\begin{eqnarray}
{\cal R} &:& | \dots \bigl ( \bigl ( ab \bigr )_d c\bigr )_f \dots \rangle \mapsto 
|\dots \bigl ( a \bigl ( bc \bigr )_e \bigr )_f \dots \rangle \; , \label{A} \\  
\Phi &:& | \dots \bigl (ab \bigr ) \dots \rangle \mapsto | \dots \bigl ( b\, a \bigr ) \dots 
\rangle \; , \label{B} 
\end{eqnarray}
exhaust all possible transformations between pairs of binary coupling schemes for any $n$. We 
shall refer to this statement as Biedenharn-Louck theorem (topic 12 in \cite{BiLo}). 
Interpreted as transformations on binary trees, ${\cal R}$ and $\Phi$, represented 
pictorially in Fig.\ref{rotation} and Fig.\ref{twist}, are referred to as rotations and 
twists, respectively.     

\begin{center}
{\bf Fig. 2} \\ 
{\bf Fig. 3} 
\end{center}

\begin{figure}[htb]
\leavevmode
\epsfysize=6cm
\epsfbox{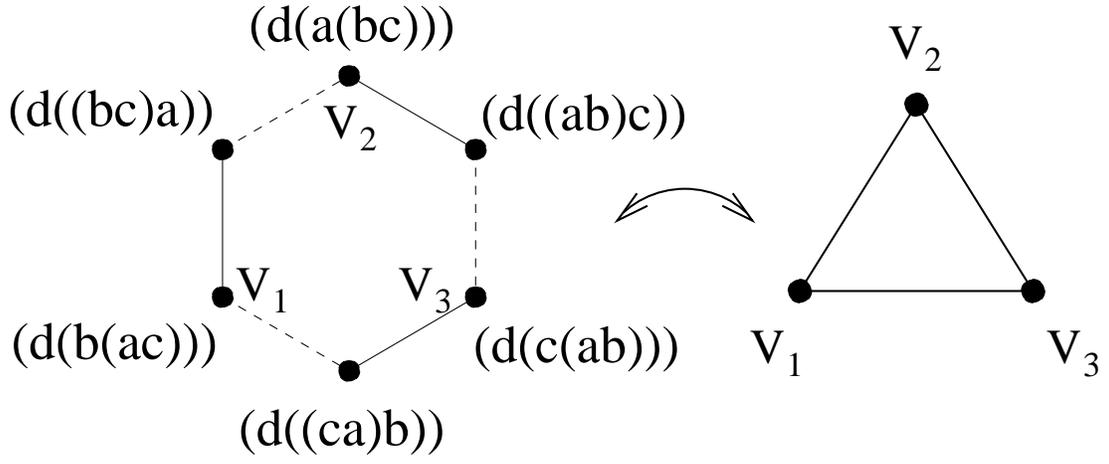}
\bigskip
\caption{On the left there appears a local configuration representing
six binary bracketings on four angular momentum variables $\{a,b,c,d\}$. 
Dotted
lines represent twist operations (phase transforms on the corresponding 
state vectors),
while the other edges are associated with rotations (Racah transforms 
between 
state vectors). On the right the reduced configuration is shown, where 
now the bonds
stands for one Racah transform plus some suitable phase/weight factors.
The graph in Fig.\ref{grafo} is built up taking into account this 
reduction procedure, 
and in particular its vertices $\{1,4,5\}$ correspond to the
vertices $\{V_1,V_2,V_3\}$ displayed here. }
\label{exagon}
\end{figure}

The coding proposed above requires both types of operations, (\ref{A}) 
and (\ref{B}), and the corresponding graph is the full twist--rotation graph 
${\mathfrak{G}}_n$, the vertices of which are to be associated with the computational 
states of the Machine and the bonds with either Racah or phase transforms. 
However, in what follows, in order to make exemplification simpler we shall limit 
our attention to {\em rotation} graphs only, {\it i.e.} ${\mathfrak{G}}_n$-graphs in 
which adjacent vertices differing only for a twist are identified, since they actually 
capture all the essential mathematical properties of our model. Fig.\ref{exagon} exhibits 
for the case $n=3$ the local reduction of ${\mathfrak{G}}_3$ when such identification 
is implemented (cfr. \cite{AqCo}).  

\begin{center}
{\bf Fig. 4} 
\end{center} 

\begin{figure}[htb]
\leavevmode
\hspace{+2cm}
\epsfbox{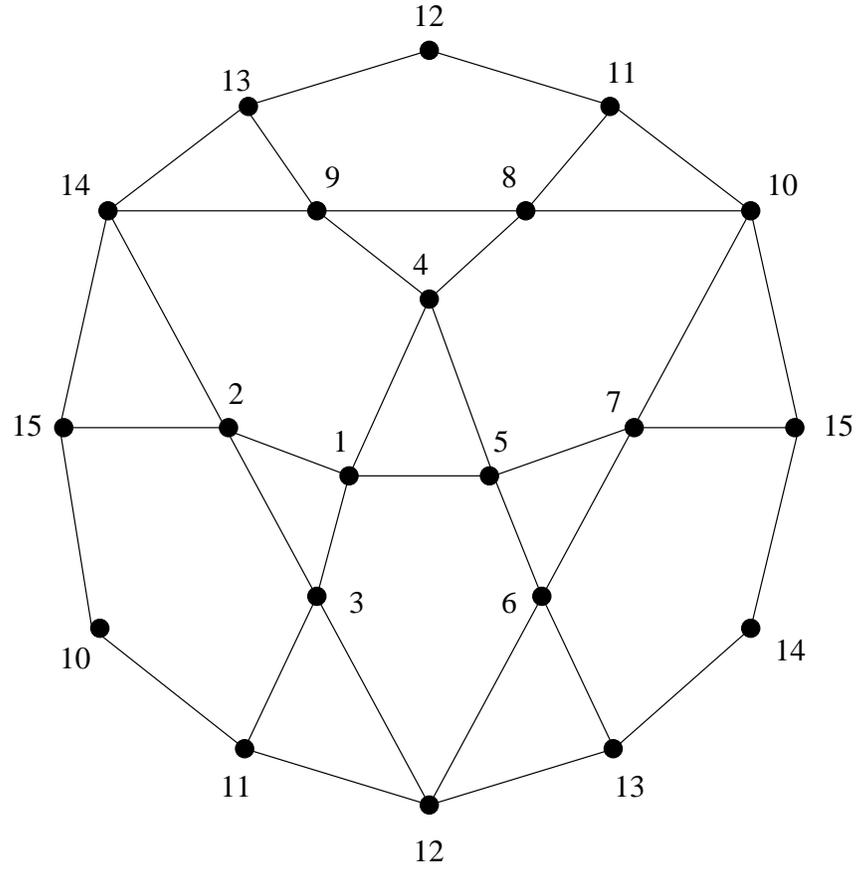}
\bigskip 
\caption{The rotation graph ${\mathfrak{G}}_3$.
Each vertex represents a binary coupling scheme of
$(n+1)=4$ angular momenta, two examples of which were given
in Fig.\ref{binary}. The picture does not exhibit
crossings but the vertices on the perimeter have to be 
identified through the action of the antipodal map
(showing that ${\mathfrak{G}}_3$ is not planar).
If we omit the intermediate labels of any binary 
bracketing of arguments $\{a,b,c,d\}$, the correspondences 
with the vertices are: 
$1\leftrightarrow (d(b(ac)))$;
$2\leftrightarrow (b(d(ca)))$;
$3\leftrightarrow ((ac)(bd))$;
$4\leftrightarrow (d(a(bc)))$;
$5\leftrightarrow (d(c(ab)))$;
$6\leftrightarrow (c(d(ab)))$;
$7\leftrightarrow ((ab)(cd))$;
$8\leftrightarrow (a(d(bc)))$;
$9\leftrightarrow ((ad)(bc))$;
$10\leftrightarrow (a(b(cd)))$;
$11\leftrightarrow (a(c(bd)))$;
$12\leftrightarrow (c(a(bd)))$;
$13\leftrightarrow (c(b(ad)))$;
$14\leftrightarrow (b(c(da)))$;
$15\leftrightarrow (b(a(cd)))$.}
\label{grafo}
\end{figure}

Accordingly, the bonds of the rotation graph (that we still denote ${\mathfrak{G}}_n$) 
correspond to Racah transforms, possibly apart from weight/phase factors. 
Fig.\ref{grafo} shows an example of such reduced computational space for $n=3$. 

\begin{center}
{\bf Fig. 5} 
\end{center} 

The combinatorial structure of ${\mathfrak{G}}_n$ is fully determined by the 
identities connecting $6j$ symbols \cite{VaMoKh}: 
\begin{description}
\item{i)} the Biedenharn-Elliot identity generates pentagon plaquettes in ${\mathfrak{G}}_n$:
$$
\sum_{x} (-)^{R+x} (2x+1) \left \{ {{a\, b\, x}\atop{c\, d\, p}} \right \} \left \{ 
{{c\, d\, x}\atop{e\, f\, q}} \right \}  \left \{ {{e\, f\, x}\atop{b\, a\, r}} \right \} 
= \biggl \{ {{p\, q\, r}\atop{e\, a\, d}} \biggr \} \left \{ {{p\, q\, r}\atop{f\, b\, c}} 
\right \} \; ,
$$
\item{ii)} Racah's identities generate triangles:
$$
\sum_{x} (-)^{p+q+x} (2x+1) \left \{ {{a\, b\, x}\atop{c\, d\, p}} \right \} \left \{ 
{{a\, b\, x}\atop{d\, c\, q}} \right \} = \biggl \{ {{a\, c\, q}\atop{b\, d\, p}} \biggr \} \; , 
$$
\end{description}
here the spin variables $\bigl \{ a, b, c, \dots , x \bigr \}$ ranges over all possible values 
in $\bigl \{ 0, \frac{1}{2}, 1, \frac{3}{2}, \dots \bigr \}$ which obey the required triangular 
conditions, and $R=a+b+c+d+e+f+p+q+r$. We argue that the greater computational power of a quantum 
computer can be ascribed to the feature that its state space 'volume' grows very rapidly. Indeed  
the order ($\#$ of vertices) of ${\mathfrak{G}}_n$ is $|{\mathfrak{G}}_n|=(2n-1)!!$ ($\sim n^n$ 
for large $n$), whereas the diameter ${\mathfrak{d}}_n$ of ${\mathfrak{G}}_n$ grows approximately 
as $n \ln n \sim \ln |{\mathfrak{G}}_n|$ \cite{FaLiVa}. In the present scheme ${\mathfrak{d}}_n$ 
is an upper bound for the time-length (number of steps) of the computations machine ${\mathfrak{M}}$ 
can perform (notice that for the full twist-rotation graph the cardinality is a factor $2^n$ larger).   

\noindent {\bf Universality} 

Universality of ${\mathfrak{M}}$, being its computational space ${\mathfrak{G}}_n$, is 
guaranteed by Biedenharn-Louck's theorem (which plays the role of a sort of generalized 
Cayley's theorem): \underline{any} unitary transformation corresponding to an operation 
of ${\mathfrak{M}}$ can be reconducted to a finite sequence of operations in ${\mathfrak{G}}_n$. 

This gives an answer to the question raised by Feynman about universality \cite{Fey}, 
explicitly defining the class of 'exact imitators' of any finite, discrete quantum system, 
with no need of resorting to the notions coming from the (inherently classical) Boolean circuit 
theory.     
 
\noindent {\bf Identification with Feynman's Q-Simulator} 

${\mathfrak{M}}$ has all the requisites of the 'quantum simulator' as defined by Feynman \cite{Fey}: 
{\em locality}, reflected in the bracketing structure, which bears on the existence of local interactions; 
{\em discreteness}, both of the computational space and of 'time', and {\em universality}. The time lapse 
from $|in\rangle$ to $|out\rangle$, as required in Feynman's scheme and as we shall see in the next section, 
is simulated both through the ordering induced by the graph combinatorial structure and by the number 
of computational steps; in other words, it is not only discrete but intrinsically inherent to the simulator  
structure \cite{Ll}. It is the interplay between such space and time discreteness which gives rise in a 
natural way to entanglement, due to the clustering proper to the (non associative) Hilbert space tensor 
product structure generated by the recoupling.  

\bigskip 
\section{Dynamics and Programming} 

The above ingredients completely define the kinematical structure of ${\mathfrak{M}}$. 
Further notions are necessary to equip it with the dynamical structure necessary to 
make it operate. 

As in a classical Turing Machine \cite{Man}, computation is a map from the {\sl input} data to the 
{\sl output} state: ${\cal U}_{\cal P} : |in\rangle \limo |out\rangle$, where now the machine states are 
coded in vectors of the Hilbert spaces ${\cal H}_n^j$ corresponding to the vertices of ${\mathfrak{G}}_n$ 
and ${\cal U}_{\cal P}$ is the class of unitary tranformations induced by the program ${\cal P}$ and 
defined by the corresponding $3nj$ symbols.  
 
The structure of computation in ${\mathfrak{M}}$ is a generalization of the conventional Boolean scheme. 
To begin with, the coding language is based on an $'$alphabet$'$ consisting in all the (possibly different)  
values $j_{\ell} \; ,\; \ell = 1,\dots ,n+1$, of the coupled angular momenta, the intermediate variables 
$k_1,\dots ,k_{n-1}$, as well as the bracketing structure, and is therefore much more powerful and flexible. 

The program ${\cal P}({\mathfrak{A}})$ to perform algorithm ${\mathfrak{A}}$ is the specification of a 
suitably designed ({\it i.e.} depending on ${\mathfrak{A}}$) \underline{ordered} sequence of $'$local 
alterations$'$ of the alphabet elements in the running state, which play the role of gates. Such alterations    
are transforms of type (\ref{A}), possibly accompanied by local permutations of labels and/or subtrees 
(moves of type (\ref{B}), phase swaps) in ${\mathfrak{G}}_n$. We shall denote any such sequence by 
$|{\mathfrak{b}}_{\alpha}\rangle$, where index $\alpha$, which keeps track of the given ordering, is such 
that $|{\mathfrak{b}}_{\alpha +1}\rangle$ is connected to $|{\mathfrak{b}}_{\alpha}\rangle$ by the elementary 
move corresponding to the local operation required by ${\mathfrak{A}}$, while $|{\mathfrak{b}}_{0}\rangle \equiv  
|in\rangle$ and $|{\mathfrak{b}}_{L}\rangle \equiv |out\rangle$. $L=L({\cal P}({\mathfrak{A}}))$ is the number 
of elementary steps required by program ${\cal P}({\mathfrak{A}})$ to complete algorithm ${\mathfrak{A}}$. 
The associated lexicographically ordered sequence $\{ |{\mathfrak{b}}_{\alpha}\rangle \, |\, \alpha = 0,\dots ,L \}$ 
defines a directed path in ${\mathfrak{G}}_n$ of length $L$ in one-to-one correspondence with the duration 
of ${\cal P}$ in units of its intrinsic discrete time step $\tau$.  

The associated unitary ${\cal U}_{{\cal P}({\mathfrak{A}})}$ 
\begin{eqnarray}
 \langle out |\, {\cal U}_{{\cal P}({\mathfrak{A}})} | in \rangle = \; : \prod_{\alpha =0}^{L-1} 
\langle {\mathfrak{b}}_{\alpha +1} |\, {\cal U}_{{\cal P}({\mathfrak{A}})} | {\mathfrak{b}}_{\alpha} \rangle : \; , 
\label{ord-prod} 
\end{eqnarray}
where $: \bullet :$ denotes $'$ordered product$'$, is a sort of superselection rule which induces 
destructive interference of the forbidden ({\it i.e.} not leading to the correct result) paths in ${\mathfrak{G}}_n$. 
Moreover, each elementary transfer matrix in (\ref{ord-prod}) can be associated with a local hamiltonian 
operator 
\begin{eqnarray}
\langle {\mathfrak{b}}_{\alpha +1} |\, {\cal U}_{{\cal P}({\mathfrak{A}})} | {\mathfrak{b}}_{\alpha} \rangle = 
\exp \left \{ i H ( {\mathfrak{b}}_{\alpha} , {\mathfrak{b}}_{\alpha +1})\, \tau \right \} \; . \label{ham}  
\end{eqnarray} 
It is worth noticing that $'$local$'$ is here intended with respect to the computational space ${\mathfrak{G}}_n$ 
of ${\mathfrak{M}}$. When (\ref{ham}) is inserted in (\ref{ord-prod}), hamiltonians $H ( {\mathfrak{b}}_{\alpha} , 
{\mathfrak{b}}_{\alpha +1})$ generally do not commute with each other and are $'$virtual$'$, in the sense that they 
are generated by the machine dynamics in the course of the computation process. In the physical interpretation, 
however, they correspond to complex polylocal, many angular momentum binary interactions and simulate even more 
complex quantum physical systems ({\sl e.g.}, sets of interacting entangled fermions and/or bosons). 
 
\noindent {\bf Optimal computation and complexity} 
 
Given two generic states $|in\rangle$, $|out\rangle$ in ${\mathfrak{G}}_n$, one can consider the Inverse Problem,  
namely read from the minimum length path the optimal algorithm that computes $|out\rangle$ as result of the 
application of some ${\cal U}_{{\cal P}({\mathfrak{A}})}$ to $|in\rangle$. The problem of finding the minimum-length 
path between two given vertices in ${\mathfrak{G}}_n$ is a $'$hard$'$ combinatorial problem, conjectured to be 
an {\bf NP}-c problem (the question however is still open, \cite{Pal} and references therein): we argue that 
the spin network simulator ${\mathfrak{M}}$ may support algorithms to solve in polynomial time such kind of problems, 
because it is known \cite{SlTaTh} that, at least in the case of unlabelled terminal nodes, the maximum distance 
between any pair of binary trees with $N$ internal nodes is at most linear in $N$.  

\noindent {\bf Path-sum Interpretation and Topological Quantum Computation} 

The dynamical behaviour of the spin network simulator is closely related to topological quantum field theories 
\cite{Qui}. In particular, the sum over all ordered paths in ${\mathfrak{G}}_n$ between $|in\rangle$ and $|out\rangle$ 
of the functionals introduced in (\ref{ord-prod}) has the form of a path-integral for a discrete topological quantum 
theory in (0+1) space-time dimensions. The topological (combinatorial) invariance is ensured by the equivalence of
paths in ${\mathfrak{G}}_n$ under the set of identities for the $6j$ symbols discussed in Sect. II. Discretized models 
based on the recoupling theory of angular momenta have been extensively studied also in 3- and 4-dimensional quantum 
gravity on the grounds of the seminal paper \cite{PoRe} (see, {\sl e.g.} \cite{CaMa} and references therein). 

On the other hand, continuous gravity in (2+1) dimensions is well described by a Chern-Simons-Witten topological quantum 
field theory, whose basic objects are closed surfaces $\Sigma$ of genus $g$. Freedman {\it et al.} in \cite{FrLaWa} 
resort just to the latter theory, by considering 'unitary topological modular functors' $h$, {\sl i.e.} operations which 
-- assigned a finite dimensional Hilbert space ${\cal H}(\Sigma )$ to any such surface -- connect diffeomorphic pairs 
$\Sigma$, $\Sigma '$. To each $h$ there corresponds a transformation ${\cal H}(\Sigma ) \longrightarrow {\cal H}(\Sigma 
')$, realized in a quantum algebra. For special choices of this algebra, any such transformation is shown to be in 
one-to-one correspondence with a product of $\nu$ elementary gate-operations, with ${\cal H}$ interpreted as the 
computational space of a Quantum Circuit Model. The interger $\nu$, which measures the complexity of the corresponding 
'computation', is linear in the length $\lambda$ of $h$ as a word in the standard generators (Dehn's twists) of the 
'mapping class group' of $\Sigma$ (whose compositions are cobordisms). Since $\lambda$ is in turn linear in the genus 
$g$, Freedman concludes that there exists a quantum circuit model that can efficiently simulate any topological modular 
functor in the given class.  On the other hand, as the presentation of a 3-manifold by surgery and triangulations are 
equivalent \cite{Rob}, the approach described above can be in principle reconducted to a subclass of spin network 
simulators. This, roughly speaking, bears on the property that the cobordisms on a continuous manifold characterizing 
the modular functor can be translated into combinatorial moves over triangulations.     

\bigskip 
\section{Conclusions and further developments}

We have exploited the kinematics and dynamics of a universal quantum simulator which encodes information in the 
full structure of the binary coupling schemes of angular momenta and manipulates such information by recoupling.  
The proposed model naturally exhibits all the characteristic features of the conventional model for quantum 
computation, such as entanglement, intrinsic parallelism, tensor product structure of the state space, as well 
as the set of requirements proposed by Feynman as essential for the correct description of a quantum simulator. 

The model raises a number of intriguing questions, which of course demand extensive consideration; we mention 
just a couple of examples: 

\smallskip 
\noindent {\bf Semiclassical Limit of the Simulator}

Given $n$, if the spin variables $j_1,\dots ,j_{n+1},j$, together with the intermediate $k_1,\dots ,k_{n-1}$, 
are all $\gg 1$ (in $\hbar$ units), each $6j$ symbol representing a Racah transformation can be approximated according 
to the asymptotics established in \cite{PoRe}
\begin{eqnarray}
\left \{ {{j_1\, j_2\, k_1}\atop{j_3\, j\, k_2}} \right \} \doteq \left \{ {{j_1\, j_2\, j_5}\atop{j_3\, j_4\, j_6}} 
\right \} \sim \sqrt{\frac{1}{12\pi V(T)}} \, \exp \left \{ i\left ( \sum_{r=1}^6 \ell_r \theta_r + 
{\scriptstyle{\frac{1}{4}}}\pi \right ) \right \} \; , \nonumber 
\end{eqnarray} 
where $V(T)$ is the euclidean volume of the tetrahedron $T$ spanned by the six edges of $'$length$'$ $\ell_r = j_r + 
\frac{1}{2}$, and $\theta_r$ is the angle between the outward normals to the faces which share $\ell_r$ (in the 
classical context canonically conjugate to $\ell_r$).   

This opens the intriguing possibility of bridging the quantum Turing Machine model to a classical 
counterpart and hence of interpreting in terms of classical actions the algorithms that it can solve.  
  
\noindent {\bf Mixed States Computation}

In principle, the whole conceptual scheme above can be reformulated in terms of density matrix formalism ({\sl i.e.} 
resorting no longer to sharp eigenstates of the $j_{\ell}$'s but rather to generalized multipole moments \cite{LoBi}). 
For example, in the simple case of two systems characterized by quantum numbers $(\sigma_i), j_i, m_i \, , \, i=1,2$, 
where $(\sigma_i)$ denotes all quantum numbers that are distinct from angular momenta eigenvalues, consider the Wigner 
coupling $|(\sigma_1 ) j_1 m_1 \rangle \otimes |(\sigma_2 ) j_2 m_2 \rangle \longrightarrow |(\sigma_1 \sigma_2 j_1 j_2 ) 
j m \rangle$. For given expansion of each single density matrix $\rho_i$, $i=1,2$,  
\begin{eqnarray}
\langle (\sigma'_i)j'_i m'_i |\, \rho_i | (\sigma_i)j_i m_i \rangle = \sum_{k_i,\kappa_i} \bigl (\sigma'_i j'_i |\!|\, 
\rho_i |\!|\sigma_i j_i \bigr )_{\kappa_i}^{k_i}\, C_{m_i \kappa_i m'_i}^{j_i k_i j'_i} \; , \nonumber  
\end{eqnarray}  
in terms of the reduced matrix elements and of the usual Clebsch-Gordan coefficients, one gets the expansion of the 
tensor product density matrix for the coupled system  
\begin{eqnarray} 
\langle (\sigma'_1 \sigma'_2 j'_1 j'_2 ) j' m' | \rho_1 \otimes \rho_2 |(\sigma_1 \sigma_2 j_1 j_2 ) j m \rangle 
= \sum_{k \kappa} ( \sigma'_1 \sigma'_2 j'_1 j'_2 j' |\!| \rho_1 \otimes \rho_2 |\!| \sigma_1 \sigma_2 j_1 j_2 j) 
\, C_{m \kappa m'}^{j k j'} \; . \nonumber 
\end{eqnarray}
By standard methods of tensor operator theory \cite{LoBi}, the reduced matrix elements above read  
\begin{eqnarray} 
( \sigma'_1 \sigma'_2 j'_1 j'_2 j' |\!| \rho_1 \otimes \rho_2 |\!| \sigma_1 \sigma_2 j_1 j_2 j) = 
\sum_{k_1 k_2 \kappa_1 \kappa_2} W_{j k j'_1 j'_2} \left \{ \begin{array}{c c c} j_1 & j_2 & j \\ k_1 & k_2 & k \\ 
j'_1 & j'_2 & j' \end{array} \right \} 
\bigl (\sigma'_1 j'_1 |\!|\, \rho_1 |\!|\sigma_1 j_1 \bigr )_{\kappa_1}^{k_1}\, \bigl (\sigma'_2 j'_2 |\!|\, \rho_2 
|\!|\sigma_2 j_2 \bigr )_{\kappa_2}^{k_2}\,  C_{\kappa_1 \kappa_2 \kappa}^{k_1 k_2 k} \; , \nonumber 
\end{eqnarray} 
where $W_{j k j'_1 j'_2}= \bigl [ (2j+1) (2k+1) (2j'_1+1) (2j'_2+1) \bigr ]^{\frac{1}{2}}$, and the recoupling 
coefficients entering are $9j$ symbols. This complicated expression gives the most general formula 
needed to describe any desired coupling by a sequence of binary couplings of density matrices, as required {\sl 
e.g.} in a realistic quantum circuit implementation.  

We expect that the framework provided by the above remarks may permit including the environment in the picture, 
for example describing the simulator by a systems of pure angular momenta and the environment coupled one to 
another, either in terms of a density matrix or in the semiclassical approximation. 


\vfill
\newpage 

\vfill
\newpage

\vfill
\newpage

\vfill
\newpage

\vfill
\newpage

\end{document}